# Hysteresis-Free Nanosecond Pulsed Electrical Characterization of Top-Gated Graphene Transistors

Enrique A. Carrion, Andrey Y. Serov, Sharnali Islam, Ashkan Behnam, Akshay Malik, Feng Xiong, Massimiliano Bianchi, Roman Sordan and Eric Pop

*Abstract*—We measure top-gated graphene field effect transistors (GFETs) with nanosecond-range pulsed gate and drain voltages. Due to high-$\kappa$ dielectric or graphene imperfections, the drain current decreases ~10% over time scales of ~10 $\mu$s, consistent with charge trapping mechanisms. Pulsed operation leads to hysteresis-free *I-V* characteristics, which are studied with pulses as short as 75 ns and 150 ns at the drain and gate, respectively. The pulsed operation enables reliable extraction of GFET intrinsic transconductance and mobility values independent of sweep direction, which are up to a factor of two higher than those obtained from simple DC characterization. We also observe drain-bias-induced charge trapping effects at lateral fields greater than 0.1 V/$\mu$m. In addition, using modeling and capacitance-voltage measurements we extract charge trap densities up to $10^{12}$ cm$^{-2}$ in the top gate dielectric (here Al$_2$O$_3$). Our study illustrates important time- and field-dependent imperfections of top-gated GFETs with high-$\kappa$ dielectrics, which must be carefully considered for future developments of this technology.

## I. INTRODUCTION

GRAPHENE DEVICES are promising candidates for nanoelectronics [1, 2] due to good electrical properties, such as high mobility [3] and high saturation velocity [4]. Other than carbon nanotubes (CNTs), graphene is the only material with intrinsic electron and hole mobilities that are both high (10,000 cm$^2$V$^{-1}$s$^{-1}$or higher) at room temperature and equal, due to symmetric conduction and valence energy bands. In contrast, transistor materials from Si and Ge to III-V compounds have reasonably good electron mobility (up to 800 cm$^2$V$^{-1}$s$^{-1}$ for Si electron inversion layers and 30,000 cm$^2$V$^{-1}$s$^{-1}$ for InAs and InSb quantum wells), but hole mobility two to twenty times lower (200 cm$^2$V$^{-1}$s$^{-1}$ for Si hole inversion layers and up to 1000 cm$^2$V$^{-1}$s$^{-1}$ for strained InSb) [5].

In the past few years, practical circuits have been demonstrated using graphene field effect transistors (GFETs), including amplifiers [6, 7], inverters [8], ring oscillators [9], radio-frequency (RF) mixers [10-12] and wafer-scale circuits [13]. However, depending on the high-permittivity (high-$\kappa$) top gate dielectric used, the graphene-dielectric interface, and the testing conditions (e.g. air ambient vs. vacuum) GFETs often exhibit characteristics that depend on the voltage sweep direction, i.e. hysteresis. The hysteresis shift can be defined as the difference in Dirac voltage ($V_0$) between forward (FWD) and reverse (REV) gate voltage sweeps ($\Delta V_0 = V_{0,FWD} - V_{0,REV}$), where $V_0$ is the gate voltage of minimum conductivity in the graphene channel (referred to as the Dirac voltage), and can be considered analogous to the threshold voltage in traditional MOSFETs.

Hysteresis is primarily caused by charge trapping [14-16] at the graphene-dielectric interfaces and by ambient molecules (i.e. water and oxygen) in contact with the graphene surface [17]. The latter effect can be reduced or eliminated by measurements under vacuum conditions (~10$^{-5}$ Torr) [15, 18] after an annealing step [18, 19]. However, trapping at the interfaces or within the bulk of the dielectrics surrounding the graphene channel is an inherent and challenging problem. Ultimately, such trapping causes device reliability and operation issues which translate to changes in carrier concentrations, and thus introduce uncertainties when extracting parameters of interest including mobility, contact resistance, and transconductance. Similar threshold voltage instabilities had also been observed in the early years of silicon technology [20] and as recently as the last decade with the introduction of high-$\kappa$ dielectrics and metal gate stacks [21, 22]. Addressing such trapping and voltage instability issues is crucial for the continued development and accurate metrology of GFETs.

In this work, we investigate the effect of pulsed current-voltage (*I–V*) measurements on the hysteresis and extracted parameters (such as mobility) of top-gated GFETs. Sub-microsecond pulsed output characteristics of top-gated exfoliated graphene FETs [23], and micro- to millisecond pulsed transfer characteristics of back-gated FETs [14, 15, 17] were previously reported (sweep rates used in Ref. [17] range from 0.19 V/s to 4.18 V/s); the latter only probing trapping at the graphene-SiO$_2$ interface. Here we use graphene grown by large scale chemical vapor deposition (CVD) and examine the gate and drain effects of reducing drain and gate pulse widths

To appear in *IEEE Trans. Electron Dev.* DOI: 10.1109/TED.2014.2309651 (2014). This work was supported in part by Systems on Nanoscale Information fabriCs (SONIC), one of the six SRC STARnet Centers, sponsored by MARCO and DARPA, by the U.S. Air Force Office of Scientific Research (AFOSR) Young Investigator Program (YIP) grant FA9550-10-1-0082, by the National Science Foundation (NSF) grant CAREER grant ECCS-0954423, the Office of Naval Research (ONR) YIP grant N00014-10-1-0853, and by the Italian Fondazione Cariplo, grant No. 2011-0373 and Directa Plus srl.

E. Carrion, A.Y. Serov, S. Islam, A. Behnam, A. Malik, F. Xiong and E. Pop are with the Department of Electrical and Computer Engineering, Univ. Illinois at Urbana-Champaign, 208 N Wright St, Urbana, IL 61801, USA.

M. Bianchi and R. Sordan are with L-NESS, Department of Physics, Politecnico di Milano, Polo di Como, Via Anzani 42, 22100 Como, Italy.

E. Pop is now with the Department of Electrical Engineering, Stanford University, 420 Via Palou Mall, Allen Building Room 335X, Stanford, CA 94305, USA (e-mail: epop@stanford.edu).

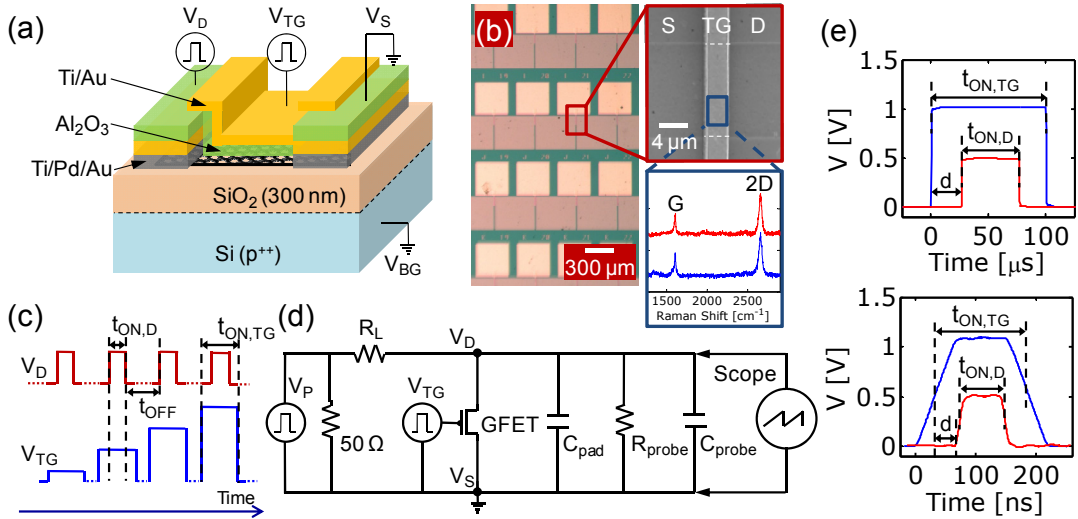

Fig. 1. (a) Schematic of top-gated graphene field-effect transistors (GFETs) fabricated in this work. (b) Optical, scanning electron microscope (SEM) images, and Raman spectra of typical devices. (c) Schematic of pulses and biases applied at top gate ($V_{TG}$) and drain ($V_D$) terminals. The amplitude of $V_{TG}$ is increased when performing a gate sweep of the voltage. (d) Diagram of circuit used to apply voltage pulses at drain and gate terminals. Current is calculated from the voltage drops across load resistor ($R_L = 0.5$-$1.5$ k$\Omega$), GFET, probe resistance ($R_{probe} = 100$ k$\Omega$) and pad ($C_{pad} = 8.3$ pF) and probe capacitances ($C_{probe} = 0.8$ pF). (e) Measured drain and top gate pulses – top gate scope connection not shown [Fig. 1(d)]– with $t_{ON,TG} = 100$ μs and $t_{ON,D} = 50$ μs (top) and $t_{ON,TG} = 150$ ns and $t_{ON,D} = 75$ ns. In both cases $V_D$ has a delay $d = t_{ON,D} / 2$ relative to $V_{TG}$. Also, $V_{TG} = 1$ V and $V_D = 0.5$ V.

down to 75 and 150 ns respectively (more than 5x smaller than the shortest pulses previously investigated [23]). We uncover two apparent trapping time constants of approximately 0.3 and 4.2 μs, ostensibly due to imperfections in the top-gate high-$\kappa$ dielectric (Al$_2$O$_3$), its interface (oxidized Al seeding layer), or the graphene itself. Hysteresis is greatly reduced when using nanosecond voltage pulses at the drain and gate terminals, effectively limiting the time over which charge trapping can occur. The extracted mobility is independent of sweep direction and up to a factor of two higher than if DC measurements were simply employed. The approach described here leads to reliable characterization of GFETs, even in the face of imperfect dielectrics and interfaces.

## II. DEVICE FABRICATION AND MEASUREMENT SETUP

Graphene is grown on copper foils similarly to our previous work [8, 24], using CVD with a methane/hydrogen mixture as precursor gases. It is then transferred onto SiO$_2$ (300 nm)/Si substrates using a dual stack of poly(methyl methacrylate) (PMMA) for support and protection (60 nm of 495 A2 and 250 nm of 950 A4). PMMA is removed using a 1:1 solution of dichloride-methane and methanol, followed by a H$_2$/Ar anneal (2 hours at 400 °C) [25]. Next, Ti/Pd/Au (0.7/20/20 nm) source/drain electrodes are fabricated using UV lithography and e-beam evaporation, followed by O$_2$-plasma channel definition and atomic layer deposition (ALD) of $t_{ox} \approx 20$ nm of Al$_2$O$_3$ (seeded by 1.5 nm of evaporated and oxidized Al). Finally, a Ti/Au (0.7/20 nm) top gate with a gate-source/drain overlap of ~150 nm is fabricated using e-beam lithography. Channel dimensions ($L$ and $W$) range from 2 – 10 μm.

Figures 1(a-b) show the schematic, optical and scanning electron microscope (SEM) images of completed devices. Raman spectra taken after transfer [inset of Fig. 1(b)] indicate that graphene is monolayer (2D-peak to G-peak integrated

intensity ratio $I_{2D}/I_G \approx 2$) and with D-peak to G-peak integrated intensity ratio $I_D/I_G \approx 0.25 \pm 0.15$ [26]. From the $I_D/I_G$ ratio we estimate [27] an average distance between Raman-active defects to be $L_a \approx 250 \pm 150$ nm. Considering micron-scale device dimensions used here, we expect the presence of defects and grain boundaries within the channel [28], and thus lower mobilities than those of exfoliated (single crystal) graphene devices [4].

During measurements, we apply voltage pulses ($V_P$) at the drain while increasing the amplitude of voltage pulses at the top gate ($V_{TG}$), as shown in Figs. 1(c-d). The $V_D$ pulse is applied after the rising ($t_R$) edge and removed before the falling ($t_F$) edges of $V_{TG}$, since gate pulse edges cause a small "resonance" on $V_D$, especially at larger amplitudes (i.e. $V_{TG} > 2$ V) and shorter edges ($t_R = t_F < 500$ ns). Hence, the full-width at half-maximum (FWHM) of $V_{TG}$ is twice that of $V_D$ ($t_{ON,TG} = 2 \cdot t_{ON,D}$) and a delay relative to $V_{TG}$ ($d = t_{ON,D} / 2$) is half the width of $V_D$ [Fig. 1(e)]. We find that these two constraints maximize signal integrity. The rise ($t_R$) and fall ($t_F$) times of gate and drain pulses vary depending on their width (i.e. $t_R = t_F = 10$ ns for $t_{ON,D} = 75$ ns and $t_R = t_F = 20 - 50$ ns for $t_{ON,TG} = 150$ ns).

The off-state relaxation time between drain pulses ($t_{OFF}$) ranges between 0.1-1 ms, which is 3-4 orders of magnitude larger than shortest $t_{ON,D}$ applied (75 ns). These off-times were found to be sufficiently long to relax all measurable effects of charge trapping from our short pulses. Larger relaxation times (up to the range of seconds) have been used while studying trapping at the graphene/SiO$_2$ [14-17] and CNT/SiO$_2$ [29] interfaces. In contrast, our analysis attempts to study and control trapping by using nanosecond-range top gate pulses ($t_{ON,TG}$) and effectively limiting the amount of carriers than can become trapped, instead of increasing de-trapping via longer off-state relaxation.

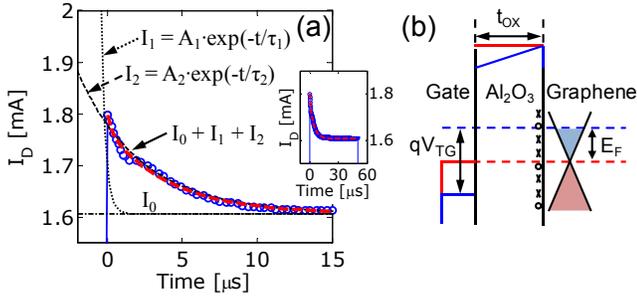

Fig. 2. (a) Measured drain current $I_D$ (blue circles and line) during the first 15 μs of a typical 50-μs pulse ($V_D = 1.9$ V, $V_{TG} = 0.5$ V). Inset shows the same data set, zoomed out for the entire 50 μs. Device measured in air ambient, $L \times W = 2 \times 10$ μm. Transient behavior is due to population of interface and bulk oxide charge traps. Current is fitted (red dashed lines) as $I_D(t) = I_0 + A_1 \cdot \exp(-t/\tau_1) + A_2 \cdot \exp(-t/\tau_2)$. Exponential terms are shifted by $I_0$ and shown in black dashed and dotted lines. Fitting parameters are $I_0 = 1.6$ mA, $A_1 = 0.03$ mA, $A_2 = 0.16$ mA, with two time constants $\tau_1 = 0.3$ μs and $\tau_2 = 4.2$ μs. (b) Schematic diagram of the metal-oxide-graphene device showing population of interface traps as the graphene Fermi level (dashed) increases from $qV_0$ (red) to $qV_{TG}$ (blue). Interface states could act as either hole (circles) or electron (x-symbols) traps.

In order to measure pulsed *I-V* characteristics we employ a pulse generator, a 1.5-GHz oscilloscope, an active probe and a simple voltage divider circuit in our setup [Fig. 1(d)]. For each top gate pulse ($V_{TG}$), a corresponding voltage pulse is applied to a load resistor ($R_L$), such that after subtracting its voltage drop ($V_{RL}$), a pulse of amplitude $V_D$ is applied to the GFET ($V_D = V_P - V_{RL}$). For $I_D$-$V_{TG}$ measurements, the amplitude of $V_D$ is kept the same throughout the measurement by adjusting the amplitude of the pulse $V_P$ at each $V_{TG}$ bias through a feedback loop (since $V_{RL}$ changes with the bias-dependent resistance of the device). Every recorded $V_D$ waveform (at a given $V_{TG}$) is an average over 200 applied pulses. The time dependence of the drain current $I_D(t)$ is obtained from the voltage drops across the load resistor ($R_L$) and a 50 Ω matching resistor (in parallel with the 50 Ω output impedance of the pulse generator), the GFET, the active probe resistance ($R_{probe}$) and pad ($C_{pad}$) and probe capacitances ($C_{probe}$), such that:

$$I_D(t) = \frac{V_P(t) - V_D(t)}{R_L + 25\,\Omega} - \frac{V_D(t)}{R_{probe}} - (C_{pad} + C_{probe}) \times \frac{dV_D(t)}{dt}. \quad (1)$$

## III. RESULTS OF PULSED MEASUREMENTS

With this setup, we first look at typical transient behavior of current [Fig. 2(a)] when $t_{ON,D} = 50$ μs pulses are applied to the drain terminal. $I_D$ reaches steady state with ~10% degradation after ~10 μs due to effect of charge traps at this particular bias condition. This drop-off is faster than previous reports [14, 17] which studied charge trapping at back gates with much thicker amorphous SiO₂ layers. The best fit of $I_D(t)$ is obtained by using two decaying exponentials of the form $A \cdot \exp(-t/\tau)$ (black dashed lines) yielding time constants $\tau_1 = 0.3$ μs and $\tau_2 = 4.2$ μs. These suggest the presence of at least two trapping mechanisms such as interface and bulk trapping [23, 30, 31]. Interface trap response times scale exponentially with their energy difference from either the valence or conduction bands of a typical channel material [30]. Since graphene does not have a band gap and trap states can be located across a wide range of energies [Fig. 2(b)], interface traps can be rapidly filled when the energy of carriers is higher than that of electron (x-symbols) or hole traps (circles). On the other hand, bulk trap response times depend on tunneling through the oxide, and thus they are expected to be slower. In our case, the oxidized Al seeding layer (AlOₓ) and graphene imperfections (i.e. grain boundaries) could be responsible for contributions to interface trapping, while the ALD-grown Al₂O₃ contributes to bulk trapping. Nevertheless, a simple tunneling front model [29] analysis reveals that such traps are likely less than ~1 nm apart in the Al₂O₃, making it challenging to ascertain their exact physical origin, which could be the topic of a future investigation.

We note that the time constants identified here (0.3 and 4.2 μs) are not originated from circuit transients, as circuit RC time constants due to $R_L$, $R_{GFET}$, $R_{probe}$, $C_{pad}$ and $C_{probe}$ are ~10 ns. However, thermal time constants of top-gated GFETs with similar geometry are of the order ~100 ns [32], thus it is possible that the shorter time constant found here ($\tau_1 = 0.3$ μs) can include a small thermal self-heating transient, which can also influence current degradation (although we note that our pulsed measurements were done at relatively low current density, ~0.1 mA/μm, except those in Section VI).

The effect of trap filling on electrical measurements can also be seen in Fig. 3(a), where the DC transfer characteristics of a typical top-gated GFET ($L \times W = 2 \times 10$ μm) show Dirac voltage shift and hysteresis ($\Delta V_0$) in air and in vacuum measurements. Charge trapping (or de-trapping) is likely to occur at the bottom graphene/SiO$_2$ interface when we vary $V_{TG}$ as the voltage drop between the graphene and back-gate is small ($V_{BG} = 0$ V). The presence of hysteresis in both air and vacuum suggests that ambient adsorbates (i.e. O$_2$, H$_2$O) and the top dielectric trapping (interface and bulk) contribute to the change in carrier density in the channel while the DC top-gate voltage is swept [14, 15]. Thus, in order to minimize such $V_0$ instabilities, we perform pulsed measurements as described above. For each $V_{TG}$ bias $I_D$ is calculated as a function of time using eq. (1) and its amplitude is averaged over the duration ($t_{ON,D}$) of each drain pulse. Figure 3(b) displays the in-air transfer characteristics for different $V_{TG}$ on-times ($t_{ON,TG}$) and compares them with simple DC $I$-$V$s using the same bias conditions ($V_D = 0.5$ V and $V_{BG} = 0$ V). As $t_{ON,TG}$ is decreased from 100 μs to 400 ns, forward (FWD) and reverse (REV) sweeps collapse onto one another and hysteresis $\Delta V_0$ disappears. In Fig. 3(c), the corresponding $\Delta V_0$ is displayed as a function of $t_{ON,TG}$ down to 150 ns; we note that for DC $I$-$V$s $\Delta V_0 = 2.3$ V, while for $t_{ON,TG} < 500$ ns hysteresis $\Delta V_0$ approaches 0 V. This $\Delta V_0$ reduction was observed across 20 devices ($L$, $W = 2$–10 μm) [Fig. 3(d)] for five testing conditions: $t_{ON,TG} = 0.15, 1, 10, 100$ μs and DC. We attribute the broadening of each distribution (corresponding to each $t_{ON,TG}$ case) to device-to-device variations, i.e. graphene quality and contact resistance.

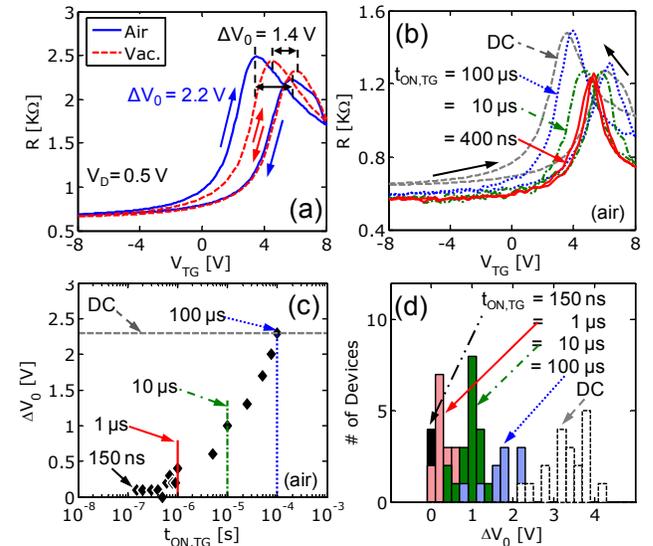

Fig. 3. (a) Hysteresis in DC measurement of resistance ($R$) vs. top gate voltage ($V_{TG}$) of a typical device ($L \times W = 4 \times 8$ μm) in air (blue-solid) and in vacuum (red-dashed). Arrows indicate sweep directions (from -8 V to +8 V and back). Hysteresis is $\Delta V_0 = 2.2$ V and 1.4 V in air and in vacuum, respectively ($V_D = 0.5$ V). (b) Typical *R-V*$_{TG}$ characteristics of another device measured in air under DC (dashed lines) and pulsed conditions. Note the suppression of hysteresis between FWD and REV sweeps as $t_{ON,TG}$ decreases to 400 ns. $L \times W = 2 \times 10$ μm. (c) Measured shift in Dirac voltage ($\Delta V_0$) from Fig. 3(b) as a function of $t_{ON,TG}$. $\Delta V_0$ is marked (lines and arrows) at five selected testing conditions: $t_{ON,TG} = 0.15, 1, 10, 100$ μs and DC ($V_{TG} = 0.5$ V). (d) Histogram of $\Delta V_0$ for 20 devices measured at same five testing conditions: $t_{ON,TG} = 0.15, 1, 10, 100$ μs and DC. Note that not all devices were tested for each case.

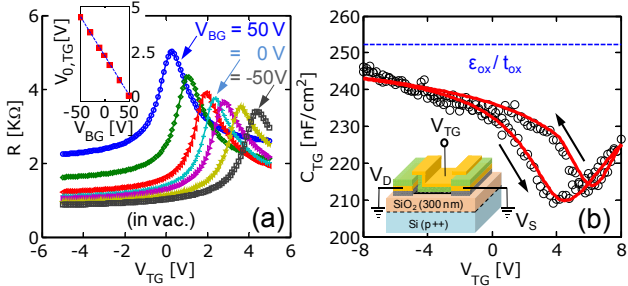

Fig. 4. (a) DC $R$-$V_{TG}$ for different $V_{BG}$ values. $L \times W = 5 \times 5$ μm, $V_D = 0.1$ V only forward sweep shown. (Inset) $V_{0,TG}$ vs. $V_{BG}$. Slope represents ratio between top and back-gate oxide capacitances ($C_{ox}/C_{BG} \approx 22$). (b) Measured top-gate capacitance at 100 kHz ($C_{TG}$, circles) and calculated (red solid line) using model described in text. Expected value from simpler extraction in (a) is also shown (dashed line). Inset shows schematic of measurement. $V_{TG}$ is applied to top-gate, source/drain are grounded and back gate is left disconnected. Side-wall and overlap capacitances which appear in parallel with $C_{TG}$ were measured in similar FET structures without graphene, and subtracted from the result.

The transfer characteristics shown in Fig. 3(b) are consistent with the presence of negative charges in the oxide. The fixed negative charges can be present in Al$_2$O$_3$ imperfections [33], and are apparent since $V_0 > 0$ V for all measurements (DC and pulsed) and sweep directions (FWD and REV). The occupied trapped states, responsible for Dirac voltage shift ($\Delta V_0$) and hysteresis, depend on pulse duration; shorter pulses limit the electrical stress time over which carriers can become trapped. We also observe that the (unified) Dirac voltage of the 400-ns pulsed sweep falls between that of the FWD and REV DC sweeps. These differences in $V_0$ are consistent with hole traps charging up in the oxide (making it less negative) when $V_{TG}$ is swept FWD starting in the hole region ($V_{TG} < V_0$), and with electron traps accumulating during the REV sweep in the electron region ($V_{TG} > V_0$) (making the oxide more negative). These additional trapped states in the oxide also contribute to the apparent variation of the channel resistance at $V_0$ in DC sweeps by increasing the charge puddle density (i.e. increasing impurity or minimum carrier densities in the channel) [4]. In contrast, when using short pulses ($< 1$ μs) less trapped states are disturbed and $V_0$ and $R(V_{TG} = V_0)$ remain constant independent of sweep direction.

## IV. GATE CAPACITANCE AND TRAP CHARGING EFFECTS

In order to estimate quantitatively oxide trapped charge densities responsible for hysteresis, we examine capacitance through measurements and modeling. First, we estimate the top dielectric capacitance as suggested in [34], by measuring the top-gate Dirac voltage ($V_{0,TG}$) shift as a function of $V_{BG}$, in vacuum. As shown in Fig. 4(a), this yields the ratio between the top- and back-gate oxide capacitance, $C_{ox}/C_{BG} \approx 22$ which gives $C_{ox} \approx 250$ nF/cm$^2$ and $\varepsilon_{ox} \approx 5.7$ for our top Al$_2$O$_3$ dielectric with oxidized Al seeding layer. Next, we measure $C$-$V$ characteristics [Fig. 4(b)] by applying DC and AC voltages to the top-gate terminal with an LCR meter. Away from the Dirac voltage, $C_{TG}$ approaches the previously estimated top-layer capacitance ($C_{ox} = \varepsilon_{ox}/t_{ox}$), while near $V_0$ it decreases and exhibits hysteresis similar to that seen in the $I$-$V$ measurements, Fig. 3. ($C$-$V$ and $I$-$V$ measurements used similar voltage sweep rates, $\sim$1.6 V/s.)

In order to estimate trapped charge densities in the top gate dielectric quantitatively, we fit a $C$-$V$ model by applying Gauss' law to our structure:

$$V_{TG} - \left[ V_0 + \frac{Q_{it}(E_F = 0)}{C_{TG}} \right] = -\frac{\left[ Q_n(E_F) + Q_{it}(E_F) \right]}{C_{TG}} + \frac{E_F}{q} \quad (2)$$

where $E_F$ is the Fermi level in graphene [Fig. 2(b)], $Q_n$ is the charge density in graphene and $Q_{it}$ is the sum of trapped charge accumulated at the AlO$_x$/graphene interface and Al$_2$O$_3$ bulk. Quantum capacitance ($C_q$) is included in our model explicitly over the density of states. The total capacitance ($C_{TG}$) is calculated [Fig. 4(b)] as a derivative of the total charge ($Q_{TG} = Q_n + Q_{it}$) to the gate voltage by varying $E_F$, such that,

$$C_{TG} = \left( \frac{\partial Q_{TG}}{\partial E_F} \right) \left( \frac{\partial V_{TG}}{\partial E_F} \right)^{-1}. \quad (3)$$

From this model we estimate (negatively charged) trapped charge densities of $7.2 \times 10^{11}$ cm$^{-2}$ and $10^{12}$ cm$^{-2}$ for the FWD and REV sweeps, respectively, at $E_F = 0$ eV (Dirac point).

## V. PULSED MOBILITY EXTRACTION

Next, we extract device transconductance ($g_m$) and effective hole mobilities ($\mu_h$) from pulsed and DC measurements [Fig. 3(b)]. We do so by first fitting a transport model ($R_{fit}$) [4, 35],

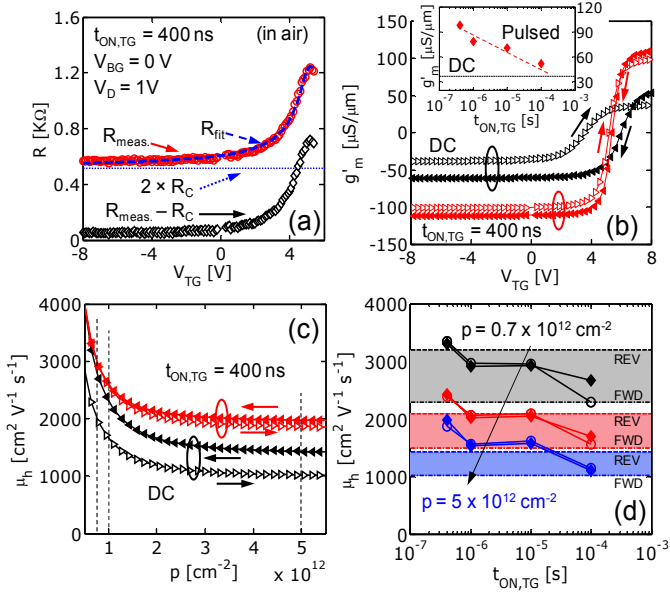

Fig. 5. (a) Hole $R$-$V_{TG}$ measured ($R_{meas}$), fitted ($R_{fit}$) as in Ref. [35] and with contact resistance ($R_C$) subtracted ($R_{meas}$–$R_C$). (b) Intrinsic transconductance ($g_m'$) from Fig. 3(b), as a function of $V_{TG}$ from pulsed (red) ($t_{ON,TG}$ = 400 ns) and DC (black) measurements. Arrows indicate direction (FWD or REV) of sweep. Inset shows maximum $|g_m'|$ (from FWD sweeps) as a function of gate on-times. (c) Extracted hole mobility as a function of carrier density ($p$) from Fig. 5(b). (d) Hole mobility vs. $t_{ON,TG}$ for different $p = 0.7$ (black), 1 (red), and $5 \times 10^{12}$ cm$^{-2}$ (blue). Values extracted from Fig. 5(c) at the marked (vertical dashed lines) densities. Open circles and solid diamonds are from FWD and REV pulsed sweeps respectively. Values extracted from DC FWD and REV sweeps mark limits of shaded regions. Note large mobility uncertainty of DC sweeps (up to ~1000 cm$^2$V$^{-1}$s$^{-1}$ or ~30 %) compared to pulsed sweeps (~50 cm$^2$V$^{-1}$s$^{-1}$ or ~2 %)

which includes contact resistance ($R_C \approx 2$–3 k$\Omega$·$\mu$m), to the measured $I_D$–$V_{TG}$ characteristics ($R_{meas}$) as shown in Fig. 5(a). Figure 5(b) then displays the *intrinsic* transconductance $g_m'$ (calculated after $R_C$ is subtracted) derived from 400-ns pulsed measurements (red) and DC measurements (black) for FWD and REV sweep directions. We note that $g_m'$ changes sign as $V_{TG}$ is swept past the Dirac point (i.e. threshold voltage) and carrier transport changes from holes to electrons. Also hysteresis is greatly reduced with 400 ns pulses compared to the DC measurement. Furthermore, the maximum value of $g_m'$ for pulsed measurements (~100 $\mu$S/$\mu$m) is approximately twice as high as the one obtained from DC measurements (~50 $\mu$S/$\mu$m). This trend is evident from the inset of Fig. 5(b), which shows the maximum $g_m'$ (from FWD sweep) as a function of gate on-time.

In Fig. 5(c) we display the effective mobility calculated as in Refs. [4, 35]. We note that mobility values are approximate since $R_C$ is fitted and not directly measured, and a constant $C_{ox}$ was used to simplify the extraction procedure. Nevertheless, this exercise illustrates the consistency and reliability of pulsed characterization vs. DC measurements. We show hole mobility ($\mu_h$) vs. carrier density for 400 ns pulses (red) and DC measurements (black), from FWD and REV sweeps. Pulsed measurements generate higher and consistent mobility values, due to reduced charge trapping. Conversely, mobility appears to be a function of sweep direction (marked with arrows) when obtained from DC $I$-$V$ measurements. We note that self-heating effects do not play a role because the mobility estimates are all done at low lateral field and low current levels, ~0.06 mA/$\mu$m, where the maximum temperature rise is at most 5 K for our GFETs [4, 32], even for the DC measurements.

Subsequently, we examine mobility dependence on $t_{ON,TG}$. Figure 5(d) shows $\mu_h$ at three carrier densities: 0.7, 1, and $5 \times 10^{12}$ cm$^{-2}$; open circles and solid diamonds represent values from FWD and REV sweeps. The mobility range from DC measurements (top and bottom lines of shaded regions) has an uncertainty up to 1000 cm$^2$V$^{-1}$s$^{-1}$ (or ~30%), while for pulsed characterization this uncertainty is significantly smaller (~50 cm$^2$V$^{-1}$s$^{-1}$ or ~2%). Once again, we note that mobility is higher at shorter pulses, due to the minimized trapped charge.

## VI. HIGH FIELD EFFECTS

We also briefly examine the effects of high lateral *intrinsic* fields $F_x'$ (after subtracting $R_C$) on $\Delta V_0$ by using our nanosecond pulsed technique. First, we find that for $F_x' \sim 0.1$ V/$\mu$m and a constant $t_{ON,TG}$ (3 $\mu$s), $\Delta V_0$ remains constant as we decrease $t_{ON,D}$ from 2 $\mu$s down to 100 ns [Fig. 6(a)]. Conversely, when we raise $F_x'$ to 0.5 V/$\mu$m, $\Delta V_0$ drastically increases as well. This increased $\Delta V_0$ caused by higher $F_x'$ occurs when hot carriers in the channel begin to fill interface or bulk trap states of the dielectric [23]. Finally, we examine $\Delta V_0$ as we decrease $t_{ON,TG}$ and replace the pulse generator at the drain terminal with a regular DC supply. We find that, at low $F_x'$, $\Delta V_0$ is equally suppressed by using a pulsed or a DC voltage at the drain terminal [Fig. 6(b)].

In general, GFETs hysteresis is a function of the amount of trapped charge at the interface and bulk of the dielectric ($Q_{it}$), which in turn affects the overall charge in the channel, capacitance and ultimately $I$-$V$ results. Also, $Q_{it}$ is a function of frequency, gate voltage ($V_{TG}$ or $V_{BG}$) and intrinsic lateral field ($F_x'$). Thus, in order to eliminate hysteresis and Dirac voltage instabilities during measurements, one should take into account these dependencies and bias devices accordingly.

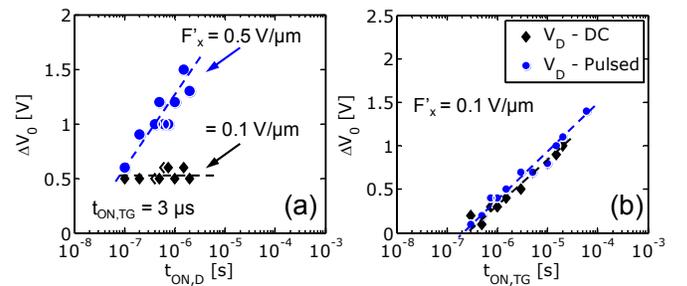

Fig. 6. (a) Dirac voltage shift ($\Delta V_0$) as a function of drain on-time ($t_{ON,D}$ = 100 ns to 2 $\mu$s and $t_{ON,TG}$ = 3 $\mu$s). The $V_{TG}$ sweeps (not shown) are from -8 V to 8 V and back to -8 V, while $V_D$ is increased from 1 V (black diamonds) to 3 V (blue circles). $V_D$ values correspond to intrinsic lateral fields $F_x' \approx 0.1$ and 0.5 V/$\mu$m, respectively, after contact resistance is subtracted ($L \times W = 3 \times 9$ $\mu$m). (b) $\Delta V_0$ vs. $t_{ON,TG}$ (300 ns to 80 $\mu$s). The drain terminal is biased using a DC (diamonds) and pulsed bias (circles) ($F_x' = 0.1$ V/$\mu$m , $t_{ON,D} = 0.5 \cdot t_{ON,TG}$).

## VII. Conclusion

In conclusion, intrinsic properties of GFETs can be probed with pulsed operation and pulses shorter than the trapping time constants of interface and bulk trapping. We also report transfer characteristics, transconductance and mobility values that do not depend on voltage sweep direction (forward or reverse) or rate. Such results "correctly" represent the *intrinsic* properties of the GFET channel, as detrimental effects from oxide and interface traps (hysteresis and $I_D$ degradation) can be greatly reduced. Finally, we show that high lateral fields can affect hysteresis and charge trapping through hot carrier injection, a situation that can also be mitigated by using short drain on-times. All of these findings shed light on careful ways to characterize graphene devices and reduce detrimental effects by using pulsed measurements, which is important for future advancement of graphene device technology.